\documentclass[fleqn,twoside]{article}
\usepackage{espcrc2}
\usepackage{cite}
\usepackage{amsmath}
\usepackage{eufrak}

\newcommand{\Li}{{\rm Li}}
\newcommand{\Sf}{{\rm S}_{1,2}}
\newcommand{\si}{{\rm sign}}

\newcommand{\bq}{\begin{equation}}  
\newcommand{\eq}{\end{equation}}
\newcommand\beq{\begin{equation}}   
\newcommand\eeq{\end{equation}}
\newcommand\bea{\begin{eqnarray}}
\newcommand\eea{\end{eqnarray}}

\newcommand\SH{\,\mbox{$\sqcup \! \sqcup$}\,}

\usepackage{graphicx}
\usepackage[figuresright]{rotating}

\title{{\small
\begin{flushleft}
DESY 04-114 \hfill {\tt hep-ph/0407044} \\
June 2004 \hfill   SFB/CPP--04--23 
\end{flushleft}}
Mathematical Structure of Anomalous Dimensions and QCD Wilson 
Coefficients in Higher Order}

\author{J. Bl\"umlein\address{Deutsches Elektronen
Synchrotron, DESY, Platanenallee 6, D-15738 Zeuthen, Germany}}%
\begin{document}

\begin{abstract}
\noindent
The alternating and non-alternating harmonic sums and other algebraic objects 
of the same equivalence class are connected by algebraic relations which 
are induced by the product of these quantities and which depend on their 
index class rather than on their value. We show how to find a basis of the 
associated algebra. The length of the basis $l$ is found to be $\leq 1/d$, 
where $d$ is the depth of the sums considered and is given by the 2nd 
{\sc Witt} formula. It can be also determined  counting the {\sc Lyndon} 
words of the respective index set. There are two further classes of 
relations: structural relations between {\sc Nielsen}--type integrals and
relations due to the specific structure of {\sc Feynman} diagrams which
lead to a considerable reduction of the set of basic functions. The relations 
derived can be used to simplify results of higher order calculations in QED 
and QCD. We also report on results calculating the 16th non--singlet moment of 
unpolarized structure functions at 3--loop order in the $\overline{\rm MS}$ 
scheme.
\vspace{1pc}
\end{abstract}

\maketitle

\section{INTRODUCTION}

\noindent
Multiple nested alternating and non-alternating harmonic 
sums $S_{a_1, \ldots, a_n}(N)$ \cite{HS1,HS2,HS3} 
emerge in perturbative higher order calculations within QED and QCD
for massless fermions,
\begin{eqnarray}
\label{eqHS}
S_{a_1, \ldots, a_n}(N) &=& \sum_{k_1 = 1}^N \sum_{k_2 = 1}^{k_1} \ldots
\sum_{k_n = 1}^{k_{n-1}} \frac{\si(a_1)^{k_1}}{k_1^{|a_1|}}
\nonumber\\ & & \ldots
\frac{\si(a_n)^{k_n}}{k_n^{|a_n|}}~.
\end{eqnarray}
Here, $a_k$ are positive or negative integers and $N$ is a positive
even or odd integer depending on the observable under consideration. One
calls $n$ the {\sf depth} and $\sum_{k = 1}^n
|a_k|$ the {\sf weight} of a harmonic sum. Harmonic sums are associated to
{\sc Mellin} transforms of real functions or {\sc Schwartz}--distributions
$f(x)$~$\epsilon~{\cal S}'[0,1]$~\cite{DISTR}
\begin{eqnarray}
S_{a_1, \ldots, a_n}(N) = \int_0^1 dx~x^N~f_{a_1, \ldots, a_n}(x)
\end{eqnarray}
which emerge in field theoretic calculations. Finite harmonic sums are
related to harmonic polylogarithms $H_{b_1, \ldots, b_n}(x)$~\cite{VR}.
Their $1/(1\pm x)$--weighted {\sc Mellin} transform yields harmonic sums.
The inverse {\sc Mellin} transform relates the harmonic sums to functions 
of {\sc Nielsen} integrals~\cite{NIELS}
of the variable $x$ at least for all sums of weight $w \leq 4$ as shown in      
\cite{HS3}, and associated generalizations for higher weight.
{\sc Nielsen} integrals are a generalization of the usual
polylogarithms~\cite{POLYL}. In the limit $N \rightarrow \infty$ the
convergent multiple harmonic sums, i.e. those where $a_1 \neq 1$, yield
(multiple) Zeta--values $\zeta_{a_1, \ldots, a_n}$, which are also called
{\sc  Euler--Zagier} sums~\cite{EZ}. A generalization of both
harmonic
polylogarithms and the {\sc Euler--Zagier} sums are the nested
$Z$--sums~\cite{MUW}, which form a {\sc Hopf} algebra~\cite{HOPF,KAS}  
and are related to {\sc Goncharov's} multiple
polylogarithms~\cite{GON}.
For a recent review see \cite{WALDS}.

Higher order calculations in massless field theories are either performed 
in {\sc Mellin}--$N$ space referring to harmonic sums or in the space of 
the momentum fractions $x$ representing the results in terms of {\sc 
Nielsen}--type integrals. The principal complexity is determined by the
amount of possible terms contributing. In the case of the 2--loop 
coefficient functions in momentum fraction space \cite{ZN} 77 different 
functions occurred, cf. 
\cite{HS3}. This number compares to the amount of all possible different  
nested harmonic sums up to weight {$w=4$}, $80 = 3^w-1$. For the 3--loop 
anomalous dimensions~\cite{L3} one expects the contribution of a wide 
class of the
$w=5$ harmonic sums and for the 3--loop coefficient functions of the $w=6$ 
harmonic sums, which means 242 or 728 sums, respectively. These sums are
not independent but are connected by different kind of relations. In the 
present paper we summarize a first class of relations recently being 
discussed in Ref.~\cite{ALGEBRA}, the so-called algebraic 
relations. It 
turns out that these relations emerge from the index-structure and the 
multiplication relation of the objects considered and are widely 
independent of other properties of the harmonic sums. In this way an 
equivalence class of even more objects is defined having the same 
properties or can be found as special cases thereof. One example is the
set of the harmonic polylogarithms~\cite{VR}. 

To obtain manageable expressions it is of importance to apply all these 
relations through which the number of basic functions to be referred to
is considerably reduced. A further reduction of the number of basic
functions follows due to structural relations~\cite{BLU1} and the
symmetry of {\sc Feynman} amplitudes.  

Experience shows that the {\sc Mellin} space 
representation yields simpler expressions in general~\cite{BM1}, which 
would not be seen easily working in $x$ space. 
To obtain as simple as possible expressions 
is of special importance because of the fact 
that data--analyses require compact results in $x$--space, which can be 
obtained using analytic continuations for the basic sums~\cite{ANCONT} and 
performing a {\sf single} numeric {\sc Mellin} inversion~\cite{BV1} for 
the whole problem. Since the evaluation of precise analytic continuations
needs special effort any possible reduction carried out before is of help.

\section{RELATIONS BETWEEN HARMO-- \newline
NIC SUMS}

\noindent
There are three types of relations between harmonic sums.
\begin{enumerate}
\item Algebraic relations due to the shuffle algebra of harmonic sums
under their multiplication, cf.~\cite{ALGEBRA,HOF,HOF1}.
\item Relations implied by structural relations of {\sc Nielsen}--type
integrals~\cite{BLU1}, which are related to harmonic sums via a {\sc
Mellin}
transforms after being weighted with the denominators $1/(1 \pm x)$

\item Symmetry relations encoded in {\sc Feynman} diagrams for the
respective quantities, which 
cancel a series of harmonic sums present in individual diagrams via
algebraic relations.
\end{enumerate}

The product of two finite harmonic sums (\ref{eqHS}) yields
\small
\begin{eqnarray}
\label{eqPROD}
&&
S_{a_1, \ldots, a_n}(N) \cdot S_{b_1, \ldots, b_m}(N)
\nonumber\\ &&
 = \sum_{l_1=1}^N \frac{\si(a_1)^{l_1}}{l_1^{|a_1|}}
  S_{a_2, \ldots, a_n}(l_1)\, S_{b_1, \ldots, b_m}(l_1) \nonumber\\ & &
+ \sum_{l_2=1}^N \frac{\si(b_1)^{l_2}}{l_2^{|b_1|}}
  S_{a_1, \ldots, a_n}(l_2)\, S_{b_2, \ldots, b_m}(l_2) \nonumber\\ & &
- \sum_{l=1}^N \frac{[\si(a_1) \si(b_1)]^l}{l^{|a_1|+|b_1|}}
  S_{a_2, \ldots, a_n}(l)\, S_{b_2, \ldots, b_m}(l)~.
\nonumber\\
\end{eqnarray}
\normalsize
We introduce the {\sf shuffle product} $\SH$ of a single  and a
general finite harmonic
sum
\begin{eqnarray}
\label{eqSTAF}  
&&S_{a_1}(N) \SH S_{b_1, \ldots, b_m}(N)
= S_{a_1, b_1, \ldots, b_m}(N) 
\nonumber\\ & & + S_{b_1, a_1, b_2, \ldots, b_m}(N)
+ \ldots + S_{b_1, b_2,  \ldots, b_m, a_1}(N)
\nonumber\\
\end{eqnarray}
which is a linear combination of the sums of depth $m+1$.
The shuffle product of two harmonic sums
of depth $n$ and $m$, $S_{a_1, \ldots, a_n}(N)$ and $S_{b_1, \ldots,
b_m}(N)$, is  
the sum of all harmonic sums of depth $m+n$ in the index set of which 
$a_i$
occurs left of $a_j$ for $i < j$ and likewise for
$b_k$ and $b_l$ for $k < l$.
As an example the shuffle product of two threefold harmonic sums is given 
by 
\small
\begin{eqnarray}
\label{eqD6C}
& &S_{a_1, a_2, a_3}(N) \SH S_{a_4, a_5, a_6}(N) = \nonumber\\ 
& &~~
S_{a_1, a_2, a_3, a_4, a_5, a_6}(N)
     +S_{a_1, a_2, a_4, a_3, a_5, a_6}(N) \nonumber\\ & & 
     +S_{a_1, a_2, a_4, a_5, a_3, a_6}(N)   
     + S_{a_1, a_2, a_4, a_5, a_6, a_3}(N) \nonumber\\ & &  
     +S_{a_1, a_4, a_2, a_3, a_5, a_6}(N)  
     +S_{a_1, a_4, a_2, a_5, a_3, a_6}(N) \nonumber\\  & &
     + S_{a_1, a_4, a_2, a_5, a_6, a_3}(N)  
     +S_{a_1, a_4, a_5, a_6, a_2, a_3}(N) \nonumber \\ & & 
     +S_{a_1, a_4, a_5, a_2, a_6, a_3}(N)   
     + S_{a_1, a_4, a_5, a_2, a_3, a_6}(N) \nonumber\\ & &
     +S_{a_4, a_5, a_6, a_1, a_2, a_3}(N)
     +S_{a_4, a_5, a_1, a_6, a_2, a_3}(N) \nonumber\\ &&
     + S_{a_4, a_5, a_1, a_2, a_6, a_3}(N)
     +S_{a_4, a_5, a_1, a_2, a_3, a_6}(N) \nonumber
\end{eqnarray} \begin{eqnarray} & &
     +S_{a_4, a_1, a_5, a_6, a_2, a_3}(N) 
     + S_{a_4, a_1, a_5, a_2, a_6, a_3}(N) \nonumber\\ & &
     +S_{a_4, a_1, a_5, a_2, a_3, a_6}(N)
     +S_{a_4, a_1, a_2, a_3, a_5, a_6}(N) \nonumber\\ &&
     + S_{a_4, a_1, a_2, a_5, a_3, a_6}(N)
     +S_{a_4, a_1, a_2, a_5, a_6, a_3}(N) 
\nonumber\\
\end{eqnarray}
\normalsize
Finally one establishes a system of linear equations in which the 
linear elements of the shuffle products form the variables and a 
polynomial out of harmonic sums of lower depth forms the inhomogeneity. We 
furthermore consider all index permutations. This system contains all 
algebraic relations. In Ref.~\cite{ALGEBRA} all solutions for harmonic 
sums up to depth $d=6$ were given. This complies to the level of 
sophistication needed to reduce the corresponding relations which emerge 
for massless 3--loop coefficient functions.

\section{NUMBER OF ALGEBRAICALLY INDEPENDENT HARMONIC SUMS}

\noindent
Let us consider the index set of a harmonic sum of depth $d$. One may 
consider this set as a {\sf word} $w$ or a non--commutative product of 
{\sf 
letters} of an {\sf ordered alphabet}  ${\mathfrak A} = \{a,b,c,d, 
\ldots\}$. Any word can be decomposed into three parts
\begin{equation}
w = pxs~,
\end{equation}
a prefix $p$, a suffix $s$, and the remainder part $x$. Among all words 
$w$ 
the {\sc Lyndon} words, cf. e.g.~\cite{REUT}, are those being smaller than 
any of its suffixes.

According to a Theorem by {\sc Radford}~\cite{RADF} the shuffle algebra 
discussed above is freely generated by the {\sc Lyndon} words, i.e. the 
length of its basis is given by the number of {\sc Lyndon} words. We would 
like to count the number of {\sc Lyndon} words for index sets, where 
the same letters can emerge repeatedly. The corresponding relation is due
to {\sc Witt}~\cite{WITT} and will be called  2nd {\sc Witt} formula
\begin{eqnarray}
\label{eqWIT2}
l_n(n_1, \ldots, n_q) &=& \frac{1}{n} \sum_{d|n_i} \mu(d)
\frac{\left(\frac{n}{d}\right)!}
{\left(\frac{n_1}{d}\right)! \ldots \left(\frac{n_q}{d}\right)!},
\nonumber\\
\end{eqnarray}
with $n = \sum_{k=1}^q n_k$.
Here $\mu(d)$ denotes the {\sc M\"obius} function.
One may derive $l_n(n_1, \ldots, l_q)$ using the generating functional 
\begin{eqnarray}
\label{eqWIT2GF}
\frac{1}{1-x_1 - \ldots - x_{n_q}} =  \prod_{n = 1}^\infty
\left(\frac{1}{1-\sum_{k=1}^q x_k^{d_k}}\right)^{l_n(n_i)}
\nonumber
\end{eqnarray}
Note that (\ref{eqWIT2}) is related to the {\sc Gauss-Witt} relation 
mentioned by {\sc Hoffman}~\cite{HOF} for the number of basic multiple 
Zeta--values of weight $w$ for $\forall n_i > 0$ if all cases for fixed 
weight are summed over. An even more strict 
relation in the inclusive case has been conjectured by {\sc 
Zagier}~\cite{EZ} and {\sc 
Broadhurst} and {\sc Kreimer}~\cite{BRKR} in the case of multiple 
Zeta--values and verified up to $w=12$.

Let us come back to Eq.~(\ref{eqWIT2}). We can draw some immediate 
conclusions out of this relation. If the numbers $n_i$ have no common 
divisor larger than 1, the number of the basis elements compared to the 
number of all objects equals $1/d$, where $d$ denotes the {\sf depth} 
of the index set. In case of common divisors larger than 1 we checked that 
the basis is always shorter for all depths up to $d=10$, see 
\cite{ALGEBRA}. 

\begin{center}  
\begin{tabular}[h]{||r||r|r|l||}
\hline \hline %
\multicolumn{1}{||c||}{\sf Weight}&
\multicolumn{1}{c|}{\sf  \# Sums}&
\multicolumn{1}{c|}{\sf  \# Basic Sums}&
\multicolumn{1}{c||}{\sf Fraction  }\\
\hline\hline  
1 &   2 &   0 &    0.0    \\
2 &   8 &   1 & 0.1250 \\
3 &  26 &   7 & 0.2692 \\
4 &  80 &  23 & 0.2875 \\
5 & 242 &  69 & 0.2851 \\
6 & 728 & 183 & 0.2513 \\
\hline \hline
\end{tabular}
\renewcommand{\arraystretch}{1.0}
\end{center}

\section{OTHER RELATIONS}

\vspace{1mm}\noindent
Further to the algebraic relations {\sf structural relations} between the
generating functions of harmonic sums, the $1/(1 \pm x)$ weighted
Nielsen--type integrals, exist. These relations reduce the number of
basic
harmonic sums further~\cite{BLU1}. As an example we mention the relation
\begin{eqnarray}
\frac{1}{2} \frac{\Li_2^2(x^2)}{1-x^2} &=& \frac{\Li_2(x)}{1-x} +
\frac{\Li_2(-x)}{1-x} \nonumber\\  &+&
\frac{\Li_2(x)}{1+x} +
\frac{\Li_2(-x)}{1+x}~. 
\end{eqnarray}

The structure of Feynman diagrams is furthermore selective on the type
of harmonic sums which may occur for the final physical results. As
mentioned before, $x$--space calculations of the  two--loop {\sc Wilson}
coefficients
resulted in the occurrence of nearly as many different functions as
combinatorially possible. Studying the {\sc Mellin} transforms of these
functions in detail \cite{BM1, JBVR} one finds that a series of harmonic
sums  occurs with a partial index symmetry such that certain classes of
basic functions do not contribute. In the case of two loop {\sc Wilson}
coefficients  the function
\begin{equation}
\label{eqln}
\frac{\ln(1+x)}{1+x}~,
\end{equation}
which is mathematically irreducible, does not contribute, although it is
of weight~2 and functions up to weight~4 span the basis. Likewise other
irreducible lower--weight functions, which cannot be eliminated by
structural relations do not occur. The function (\ref{eqln}) is, however,
instrumental to build the basis spanning the set of functions required
for the 3--loop anomalous dimensions.  

\section{A QUADRATIC LAW ?}

\noindent
The final analysis of the anomalous dimensions and {\sc Wilson}
coefficients 
\cite{BM1, JBVR, JBSM2} shows that one may come to the following
representation~:
\begin{enumerate}
\item Single harmonic sums $S_{\pm k}(N)$ and their analytic
continuation $\propto \psi^{k-1}(z)$ are regarded as well--known
functions. Likewise any derivative of the analytic continuation of the
{\sc Mellin} transforms over $[0,1]$ of the basic functions introduced
subsequently are considered to be trivial.

\item The analytic continuation of the {\sc Mellin} transforms of the
basic
functions are meromorphic functions with poles at the integers bound from
above by a positive number. They can be represented by factorial series
\cite{FACTSER}.

\item The one--loop anomalous dimensions require {\sf w=1} harmonic sums,
likewise the 1--loop {\sc Wilson} coefficient are expressed by at most
{\sf w=2}
harmonic sums, which are all trivial in the above sense.

\item The two--loop anomalous dimension are represented as polynomials of
trivial functions but the {\sc Mellin} transform of {\sf one}
non--trivial
function of up to {\sf w=3}
\begin{eqnarray}
\frac{\Li_2(x)}{1+x},~~~~[1].
\end{eqnarray}

\item The two--loop {\sc Wilson} coefficients are represented as
polynomials of
the {\sc Mellin} transform of the functions mentioned and the {\sf four}
additional basic functions
\begin{eqnarray}
\frac{\Li_2(x)}{1-x},~~\frac{\Li_3(x)}{1+x},~~~\frac{\Sf(x)}{1 \pm x},
~~~~[18,27].
\end{eqnarray}

\item The three--loop anomalous dimensions are represented as polynomials 
of the {\sc Mellin} transform of the functions mentioned and the {\sf
nine}
additional basic functions 
\begin{eqnarray}
\frac{\ln(1+x)}{1+x},~~\frac{\Li_4(x)}{1 \pm x},\nonumber\\
\frac{{\rm S}_{1,3}(x)}{1 + x},~~\frac{{\rm S}_{2,2}(x)}{1 \pm x},
\nonumber\\ 
\frac{{\rm S}_{2,2}(-x) -
\Li_2^2(-x)/2}{1 \pm x},~~\frac{\Li_2^2(x)}{1+x},~~[28].
\end{eqnarray}
\end{enumerate}

In this way the number of basic functions contributing to the single scale 
quantities discussed up to weight {\sf w = 5} leads to the
following reduction from the combinatorial complexity
\begin{equation}
2 \cdot 3^{w-1}  \rightarrow \theta(w-2) \left[w - 2 \right]^2~,
\end{equation} 
cf.~\cite{JBSM2}, i.e. the number of basic elements grows only
quadratically,
rather than exponentially. This relation will be examined for the case of
three--loop coefficient functions, i.e. for {\sf w=6}, very soon. 

\section{THE 16th MOMENT OF THE NON--SINGLET THREE LOOP ANOMALOUS
DIMENSION OF \boldmath $F_1(x,Q^2)$}

\noindent
Previous fixed--moment calculations of anomalous dimensions and
{\sc Wilson}
coefficients extended to the moments $N=14$ for the non--singlet structure
function $F_{1,L}(x,Q^2)$ and $N=13$ for $xF_3(x,Q^2)$ \cite{L3}.
Weeks before the calculation of the complete result for the 3--loop
non--singlet
anomalous dimensions \cite{ANOM3} was finished~\footnote{A corresponding
publication on the 3--loop {\sc Wilson} coefficients will appear later.}
the
calculation of the 16th moment of the non--singlet anomalous dimension of
the structure function $F_1(x,Q^2)$ and the {\sc Wilson} coefficients
for
$F_{2,L}(x,Q^2)$ was started using the {\tt MINCER} algorithm
\cite{MINCER}. The result of this calculation provides an independent test of 
the complete results. The calculation was performed using several
power--PC's partially linked to a 4.2 Tbyte raid system.~\footnote{We
thank 
S. Wiesand and P. Wegner for technical assistance and U. Gensch and C.
Spiering for their support of this calculation.}  The calculational time  
for both the projections $g_{\mu\nu}$ and $P_{\mu} P_{\nu}$ of the forward
Compton amplitude amounted to $564$ CPU days. The 16th moment of the
non-singlet anomalous dimension for $F_1^{\rm NS}(x,Q^2)$ reads,
cf.~\cite{JBJV}:
{\small 
\begin{eqnarray}
\gamma^{(0)}_{16} &=& {\frac {64419601}{6126120}}\,{ C_F}
\\
\gamma^{(1)}_{16} &=&
-{\frac {1176525373840303}{112588038763200}}\,{ C_F}\,{ N_F}
\nonumber\\ 
& &+{\frac {
21546159166129889}{484994628518400}}\,{ C_F}\,{ C_A}
\nonumber\\ 
& &-{\frac {
3689024452928781382877}{459818557352009856000}}\,{{ C_F}}^{2}
\end{eqnarray}
}
{\small
\begin{eqnarray}
\gamma^{(2)}_{16} &=& \Biggl(\frac
{59290512768143}{1563722760600}\,{\zeta_3}
\nonumber\\ &-&
{\frac
{58552930270652300886778705063429867}{3451337970612452534317096673280000}}
\Biggr)
\nonumber\\ & & \cdot~{{C_F}}^{3}
\nonumber\\ &+&
\Biggl (-{\frac {15018421824060388659436559}{
579371382263532418560000}} \nonumber\\
&-&{\frac {64419601}{765765}}\,{ \zeta_3}\Biggr ){
C_F
}\,{ C_A}\,{ N_F} \nonumber
\\ &+&
\Biggl({\frac {1670423728083984207878825467}{
6488959481351563087872000}}
\nonumber\\ 
&+&
{\frac {59290512768143}{3127445521200}}\,{
\zeta_3}
\Biggr ){ C_F}\,{{ C_A}}^{2}
\nonumber\\
&-&{\frac {5559466349834573157251}{
2069183508084044352000}}\,{ C_F}\,{{ N_F}}^{2}
\nonumber \end{eqnarray} {\small \begin{eqnarray}
&+&
\Biggl
(-{\frac {
1229794646000775781127856064477}{30335885575318557435801600000}}
\nonumber\\  &-&
{\frac {
59290512768143}{1042481840400}}\,{ \zeta_3}
\Biggr )
{{ C_F}}^{2}{
C_A}
\nonumber\\ &-&
+\Biggl
({\frac
{71543599677985155342551355451}{938967886855098206346240000}}
\nonumber\\ & &
+{\frac
{64419601}{765765}}\,{ \zeta_3}\Biggr ){{ C_F}}^{2}{ N_F}
\end{eqnarray}
}
\normalsize

\noindent
and agrees with the result in
\cite{ANOM3} for $N = 16$. The corresponding moments of the {\sc Wilson}
coefficients $C_{2,L}(x,Q^2)$ are given in \cite{JBJV}.

\section{CONCLUSIONS}

\noindent
The product of finite alternating or non--alternating harmonic sums is given
by
the shuffle product of harmonic sums and polynomials of harmonic sums of 
lower
depth. These representations imply algebraic relations between the harmonic
sums. If one considers all harmonic sums associated to an index set
$\{a_1, \ldots, a_k\}$ one may express these sums by a number of basic
sums. It turns out that this number is given by the 2nd {\sc Witt}
formula which counts the
number of {\sc Lyndon} words corresponding to the respective index set.
The set of these {\sc Lyndon} words generates in this sense all harmonic
sums of this class {\sf freely}. By solving the corresponding linear
equations we
derived the explicit representation of all harmonic sums up to depth $d=6$
without specifying the indices numerically and gave all expression which
are structurally needed to express the sums up to weight $w=6$. The
counting relations for the basis of the finite harmonic sums were given
up to depth $d=10$. The relations derived hold likewise for other
mathematical objects obeying the same multiplication relation or a simpler
one, which is being contained, as that for harmonic polylogarithms. This
is due to the fact that the relations derived depend on the index set
and the multiplication relation but on no further properties of the
objects considered.

The ratio of the number of basic sums for a given index set and
the number of all sums is mainly determined by the depth $d$ rather
than the weight of the respective sums, due to the pre-factor $1/d$ in the
{\sc Witt} formula. Modifications occur due to common non-trivial divisors
of the numbers of individual indices in the set being considered. Up to
$d=10$ we showed that the fraction of basic sums is always $\leq 1/d$
compared to all sums. The use of these algebraic relations leads to a
considerable reduction in the set of functions needed to express the
results of higher order calculations in massless QED and QCD and related
subjects. Further reductions are due to structural relations and
symmetries in the set of Feynman amplitudes calculated. 
For practical applications such as the description of the QCD
scaling violation of the structure functions in deeply inelastic
scattering the harmonic sums  occurring in the {\sc Mellin} $N$ space
calculation have to be translated to $x$--space by the inverse {\sc
Mellin} transform. For this reason the respective harmonic sums have to be
analytically continued in the argument $N$ to complex values, which
requires a high effort using numerical procedures. It is therefore
recommended to use as many as possible relations between the $N$ space
objects before the last step is being performed.

The 16th moment for the parity--conserving non--singlet structure
function was calculated. It provides a test on the complete result, which
has been derived recently for the anomalous dimensions, and on the
upcoming result for the coefficient functions.

\vspace{2mm}\noindent
{\bf Acknowledgment.} This paper was supported in part by DFG
Sonderforschungsbereich Transregio 9, Computergest\"utzte Theoretische
Physik and EU grant HPRN--CT--2000--00149. I would like to thank S. Moch,
V. Ravindran, T. Riemann and J. Vermaseren for discussions.


\end{document}